%% file: main.tex
\documentclass[10pt, twocolumn]{article}
\usepackage{epsfig}

\begin{document}

\title{Automated Generation of Layout and Control\\ for Quantum Circuits}

\author{
Mark Whitney, Nemanja Isailovic, Yatish Patel and John Kubiatowicz\\
University of California, Berkeley\\
\{whitney, nemanja, yatish, kubitron\}@eecs.berkeley.edu\\
\emph{To appear in the ACM International Conference on Computing Frontiers, 2007}
}

\date{}

\maketitle

\input{abstract.tex}

\input{intro.tex}

\input{iontraps.tex}

\input{related.tex}

\input{proposedcad.tex}

\input{control.tex}

\input{grid_layout.tex}

\begin{figure*}[ht]
\begin{minipage}{\hsize}
\begin{center}
\epsfig{file=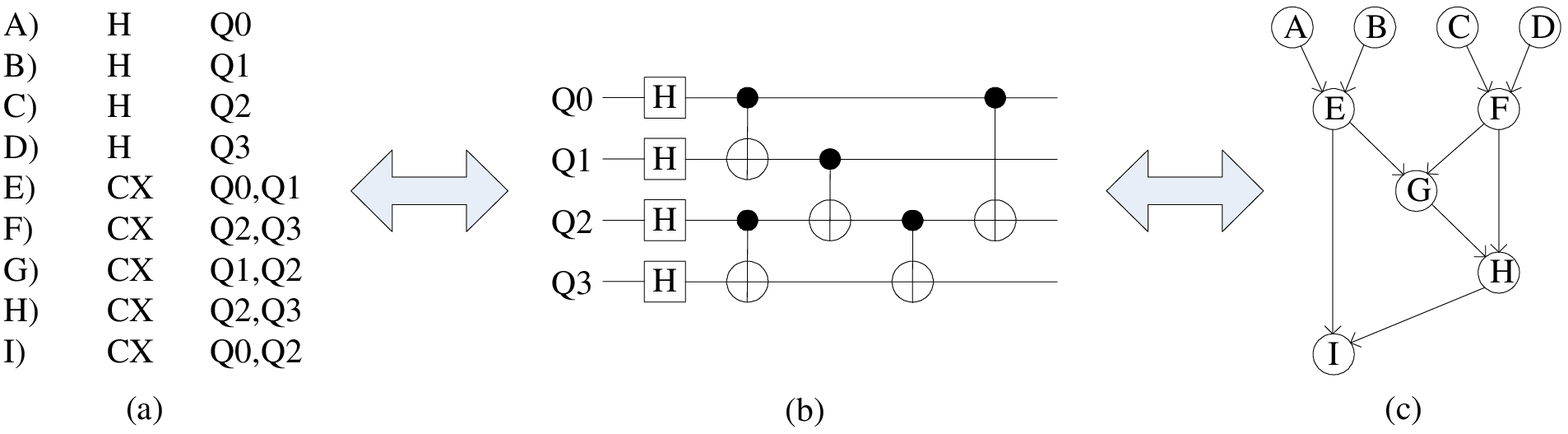, width=0.75\hsize}
\end{center}
\vspace{-20pt}
\caption{\label{fig:dataflow_graph}a) A QASM instruction sequence.
b) A quantum circuit equivalent to the instruction sequence in (a).
c) A dataflow graph equivalent to the instruction sequence in (a).  Each node
represents an instruction, as labeled in (a).  Each arc represents a qubit
dependency.}
\end{minipage}
\begin{minipage}{\hsize}
\vspace{10pt}
\begin{center}
\epsfig{file=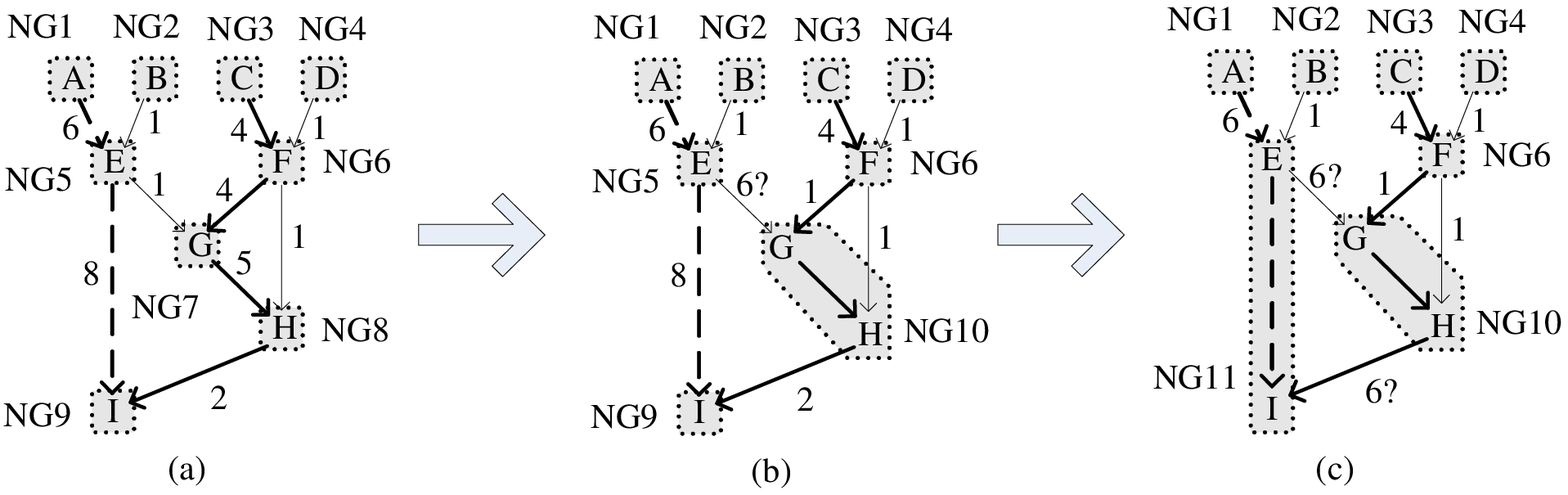, width=0.75\hsize}
\end{center}
\vspace{-20pt}
\caption{\label{fig:dataflow_alg}a) Each node (instruction) is initialized in
its own node group (NG, outlined by the dotted lines), which corresponds to a
physical gate location in a layout.  Once placed, we extract physical distances between
the nodes (the edge labels).  b) We find the longest edge weight on the longest critical
path (the length 5 edge on the path C-F-G-H-I; solid bold arrows) and merge its two node groups to
eliminate that latency.\newline c) We recompute the critical path (A-E-I; dashed bold arrows) and merge
its node groups, and so on.}
\end{minipage}
\end{figure*}

\input{greedy_layout.tex}

\input{dataflow_layout.tex}

\input{evaluation.tex}

\input{conclusion.tex}

\bibliographystyle{plain}
\bibliography{main,qubib,micro38-QuantumQLA-paper}

\end{document}

%% file: abstract.tex
\begin{abstract}

We present a computer-aided design flow for quantum circuits, complete
with automatic layout and control logic extraction.  To motivate
automated layout for quantum circuits, we investigate grid-based layouts
and show a performance variance of four times as we vary grid structure
and initial qubit placement.  We then propose two polynomial-time design
heuristics: a \emph{greedy} algorithm suitable for small, congestion-free
quantum circuits and a \emph{dataflow-based analysis} approach to placement and
routing with implicit initial placement of qubits.  Finally, we show
that our dataflow-based heuristic generates better layouts than the
state-of-the-art automated grid-based layout and scheduling mechanism in
terms of latency and potential pipelinability, but at the cost of some
area.

\end{abstract}

%% file: intro.tex
\section{Introduction}\label{sec:intro}

Quantum computing offers us the
opportunity to solve certain problems thought to be intractable on a
classical machine.
For example, the following classically hard problems benefit from
quantum algorithms:
factorization~\cite{Shor94}, unsorted database search~\cite{Grover96},
and simulation of quantum mechanical systems~\cite{zalka1998sqs}.

In addition to significant  work on quantum algorithms and
underlying physics, there  have been several studies exploring
architectural trade-offs for
quantum computers.
Most such research~\cite{balensiefer2005efa, metodi2006spo}
has focused on simulating quantum algorithms on a fixed layout rather
than on  techniques for quantum circuit synthesis and layout
generation.  
These studies tend to use hand-generated and hand-optimized layouts on which
efficient scheduling is then performed.  While this approach is quite
informative in a new field, it quickly becomes intractable as the size
of the circuit grows.

\begin{figure}
\begin{center}
\epsfig{file=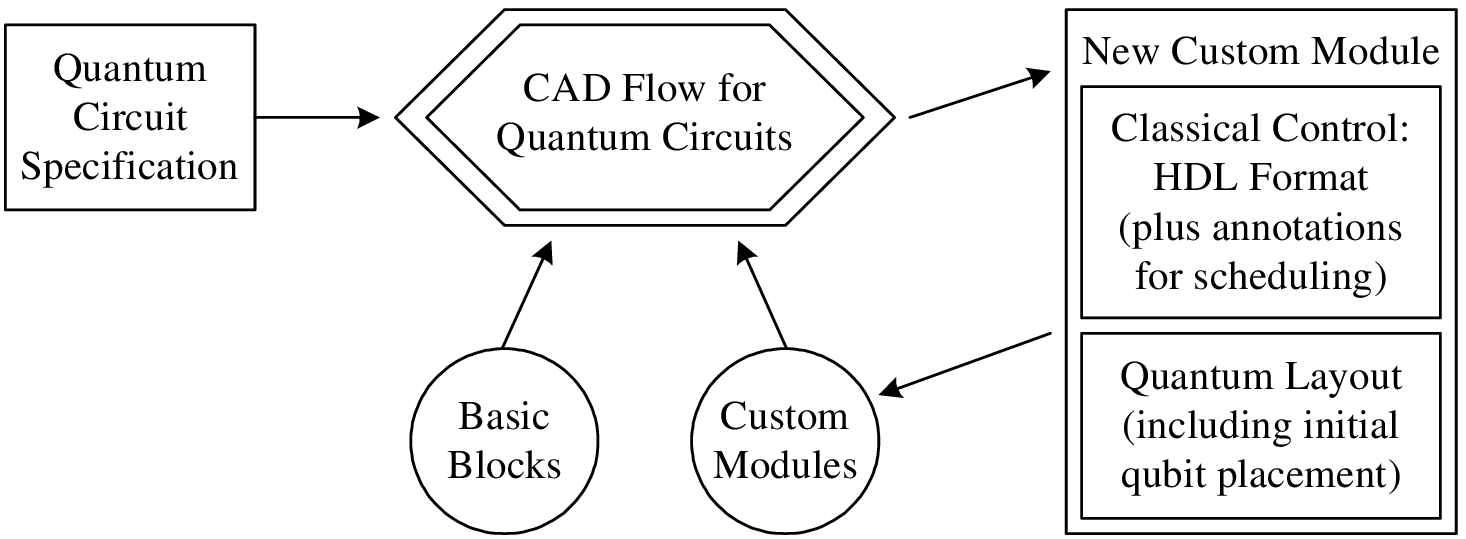,width=\hsize}
\end{center}
\vspace{-0.2in}
\caption{\label{fig:cadflowvision}The goal of our CAD flow is to automate
the laying out of a quantum circuit to generate a physical layout, an intelligent
initial placement of qubits, the associated classical control logic and annotations
to help the online scheduler better use the layout optimizations as they were
intended.  This flow may then be used recursively to design larger blocks using
previously created modules.}
\end{figure}

Our goal is to automate most of the tasks involved in generating
a physical layout and its associated control logic from a high-level quantum
circuit specification (Figure~\ref{fig:cadflowvision}).  Our computer-aided
design (CAD) flow should process a quantum circuit specification and produce
the following:
\begin{itemize}
\item a physical layout in the desired technology
\item an intelligent initial qubit placement in the layout
\item classical control circuitry specified in some
hardware description language (HDL), which may then be run through a
classical CAD flow
\item a set of annotations or ``hints'' for the online scheduler, allowing
a tighter coupling of layout optimizations to actual runtime operation
\end{itemize}

Much like a classical CAD flow, this quantum
CAD flow is intended to be used hierarchically.  We begin with a set of
technology-specific basic blocks (some ion trap technology examples are
given in Section~\ref{sec:iontraps}). We then lay out some simple
quantum circuits with the CAD flow, thus creating custom modules.
The CAD flow may then
be used recursively to create ever larger designs.  This approach allows us to
develop, evaluate and reuse design heuristics and avoids both the uncertainty
and time-intensive nature of hand-generated layouts.

\subsection{Motivation for a Quantum CAD Flow}\label{sec:quantvclass}

Quantum circuits that are large enough to be ``interesting'' require the
orchestration of hundreds of thousands of physical components. In
approaching such problems, it is important to build upon prior work in
classical CAD flows.  Although the specifics of quantum technologies
(such as are discussed in Section~\ref{sec:iontraps}) are different from
classical CMOS technologies, prior work in CAD research can give us insight
into how to approach the automated layout of quantum gates and channels.

Further, quantum circuits exhibit some interesting properties that
lend themselves to automatic synthesis and computer-aided design
techniques:

\begin{description}

\item[Quantum ECC] Quantum data is extremely fragile and consequently
  must remain encoded at all times -- while being stored, moved, and
  computed upon.  The encoded version of a circuit is often two or three
  orders of magnitude larger than the unencoded version.  Further, the
  appropriate level of encoding may need to be selected as part of the
  layout process in order to achieve an appropriate ``threshold'' of
  error-free execution. Rather than burdening the designer with the
  complexities of adding fault-tolerance to a circuit, computer-aided
  synthesis, design and verification can perform such tasks
  automatically.

\item[Ancillae] Quantum computations use many helper qubits known as
  \emph{ancillae}.  Ancillae consist of bits that are constructed,
  utilized and recycled as part of a computation.  Sometimes, 
  ancillae are explicit in a designer's view of the circuit.  Often,
  however, they should be added automatically in the process of circuit
  synthesis, such as during the construction of fault-tolerant circuits
  from high-level circuit descriptions. An automatic design flow can
  insert appropriate circuits to generate and recycle ancillae without
  involving the designer.

\item[Teleportation] Quantum circuits present two possibilities for data
  transport: \emph{ballistic movement} and \emph{teleportation}. Ballistic
  movement is relatively simple over short distances in technologies such as ion traps
  (Section~\ref{sec:iontraps}).  Teleportation is an alternative that
  utilizes a higher-overhead distribution network of entangled quantum
  bits to distribute information with lower error over longer
  distances~\cite{isailovic2006ins}. The choice to employ teleportation
  is ideally done after an initial layout has determined long
  communication paths.  Consequently, it is a natural target for a
  computer-aided design flow.
\end{description}

\subsection{Contributions}

In this paper, we make the following contributions:

\begin{itemize}
\item We propose a CAD flow for automated design of quantum circuits and detail
  the necessary components of the flow.
\item We describe a technique for automatic synthesis of the classical
  control circuitry for a given layout.
\item We show that different grid-based architectures, which have been the
  focus of most prior work in this field, exhibit vastly varying performance
  for the same circuit.
\item We present heuristics for the placement and routing of quantum
  circuits in ion trap technology.
\item We lay out some quantum error correction circuits and evaluate the
  effectiveness of the heuristics in terms of circuit area and latency.
\end{itemize}

\subsection{Paper Organization}

The rest of this paper is organized as follows.
We introduce our chosen technology in Section~\ref{sec:iontraps}, followed
by an overview of prior work in the field in Section~\ref{sec:related}.
In Section~\ref{sec:proposedcad}, we detail our proposed CAD flow and our
evaluation metrics.
In Section~\ref{sec:control}, we describe the control circuitry interface
and scheduling protocol that we use in the following sections.
Section~\ref{sec:grid_layouts} contains a study of grid-based layouts,
which have been the basis of most prior work on this subject.
In Section~\ref{sec:greedy_layout}, we present a greedy approach to
laying out quantum circuits, followed in Section~\ref{sec:dataflow_layout}
by a much more scalable dataflow analysis-based approach to layout.
Section~\ref{sec:evaluation} contains our experimental results for all
three approaches to layout generation, and we conclude in
Section~\ref{sec:conclusion}.

%% file: iontraps.tex
\section{Ion Traps}\label{sec:iontraps}

For our initial study, we choose \emph{trapped
ions}~\cite{Cirac95,MMK+95a} as our substrate technology. Trapped ions
have shown good potential for scalability~\cite{Kielpinski02}.  In this
technology, a physical qubit is an ion, and a gate is a location where a
trapped ion may be operated upon by a modulated  laser.  

The ion is both trapped and ballistically moved by applying pulse
sequences to discrete electrodes which line the edges of ion traps.
Figure~\ref{fig:macroblock_abstraction}a shows an
experimentally-demonstrated layout for a three-way
intersection~\cite{qubitturn05hensinger}.  A qubit may be held in place
at any trap region, or it may be ballistically moved between them using
the gray electrodes lining the paths.

\begin{figure}
\vspace{0.1in}
\begin{center}
\epsfig{file=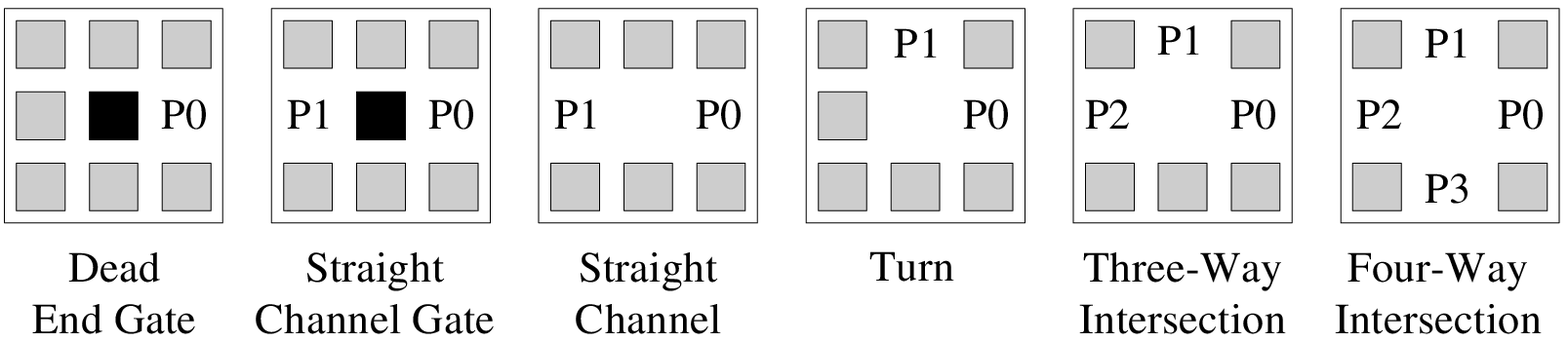,width=\hsize}
\end{center}
\vspace{-0.2in}
\caption{\label{fig:macroblocks}Example library of basic macroblocks. Each macroblock has a specific
number of ports (shown as P0-P3) along with a set of electrodes used for ion movement and trapping.
Some macroblocks contain a trap region where gates may be performed (black square).}
\end{figure}

\begin{figure*}
\begin{center}
\epsfig{file=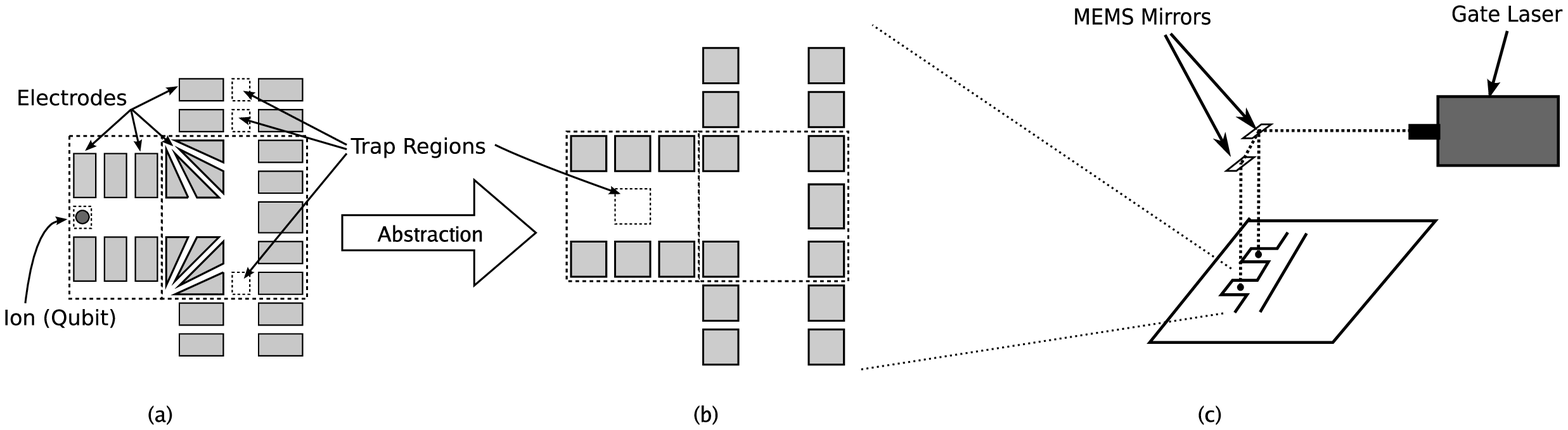,width=.8\hsize}
\end{center}
\vspace{-0.2in}
\caption{
\label{fig:macroblock_abstraction}a) Experimentally
demonstrated physical layout of a T-junction (three-way
intersection).  b) Abstraction of the
circuit in (a), built using the StraightChannel and
ThreeWayIntersection macroblocks shown in
Figure~\ref{fig:macroblocks}.
  c) The ion traps are laid out on a
plane, above which is an array of MEMS mirrors used to route and split
the laser beams that apply quantum gates.}
\end{figure*}

Rather than using ion traps as basic blocks, we define a library
of \emph{macroblocks} consisting of multiple traps for two reasons.
First, macroblocks abstract out some of the low-level details, insulating
our analyses from variations in the technology
implementations of ion traps.  Details such as which ion species is
used, specific electrode sizing and geometry (clearly variable in
the layout in
Figure~\ref{fig:macroblock_abstraction}a) and exact voltage levels
necessary for trapping and movement are all encapsulated within the
macroblock.  Second, ballistic movement along a channel requires
carefully timed application of pulse sequences to electrodes in
non-adjacent traps.  By defining basic blocks consisting of a few ion
traps, we gain the benefit that crossing an interface between basic
blocks requires communication only between the two blocks involved.

We use the library of macroblocks shown in Figure~\ref{fig:macroblocks},
each of which consists of a 3x3 grid of trap regions and electrodes,
with ports to allow qubit movement between macroblocks.  The black
squares are gate locations, which may not be performed
at intersections or turns in ion trap technology.  Each of
these macroblocks may be rotated in a layout.  This library is by no
means exhaustive, however it does provide the major pieces necessary to
construct many physical circuits.  
The macroblocks we present are abstractions of
experimentally-demonstrated ion trap technology~\cite{qubitturn05hensinger,pearson2006eip}.  In
Figure~\ref{fig:macroblock_abstraction}, we show how one can map a
demonstrated layout (Figure~\ref{fig:macroblock_abstraction}a) to our
macroblock abstractions (Figure~\ref{fig:macroblock_abstraction}b).
We model this layout as a set of StraightChannel and ThreeWayIntersection
macroblocks.
Above the ion trap plane is an array of MEMS mirrors which routes laser
pulses to the gate locations in order to apply quantum gates~\cite{kim2005sdl},
as shown in Figure~\ref{fig:macroblock_abstraction}c.

Some key differences
between this quantum circuit technology and classical CMOS are as follows:  
\begin{itemize}
\item ``Wires'' in ion traps consist of rectangular channels, lined with electrodes,
with atomic ions suspended above the channel regions and moved
ballistically~\cite{MEMStrap04madsen}.
Ballistic movement of qubits requires synchronized
  application of voltages on channel electrodes to move data around.
  Thus each wire requires movement control circuitry to handle any
  qubit communication.
\item A by-product of the synchronous nature of the qubit wire
  channels is that these circuits can be used in a synchronous manner
  with no additional overhead.  This enables some convenient
  pipelining options which will be discussed in Section~\ref{sec:graph_analysis}.
\item Each gate location will likely have the ability to perform any
  operation available in ion trap technology.  This enables the reuse
  gate locations within a quantum circuit.  
\item Scalable ion trap systems will almost certainly be two-dimensional
  due to the difficulty of fabricating and controlling
  ion traps in a third dimension~\cite{Hucul07}.  This means that all
  ion crossings must be intersections.
\item Any routing channel may be shared by multiple ions as long as
  control circuits prevent multi-ion occupancy.  Consequently, our
  circuit model resembles a general network, although scheduling the
  movement in a general networking model adds substantial complexity to
  our circuit.
\item Movement latency of ions is not only dependent on Manhattan
  distance but also on the geometry of the wire channel.
  Experimentally, it has been shown that a right angle turn takes
  substantially longer than a straight channel over the same
  distance~\cite{pearson2006eip,qubitturn05hensinger}.
\end{itemize}

%% file: related.tex
\section{Related Work}\label{sec:related}

Prior research has laid the groundwork for our quantum circuit CAD flow.
Svore et al~\cite{svore2006lsa,svore04tsa} proposed a
design flow capable of pushing a quantum program down to physical
operations.  Their work outlined various file formats and provided
initial implementations of some of the necessary tools.  Similarly,
Balensiefer et al~\cite{balensiefer2005qqa,balensiefer2005efa}
proposed a design flow and compilation techniques to address
fault-tolerance and provided some tools to evaluate simple layouts.
While our CAD flow builds upon some of these ideas, we concentrate
on automatic layout generation and control circuitry extraction.

Additionally, initial hand-optimized layouts have been proposed in the
literature.  Metodi et al~\cite{metodi2005qla} proposed a uniform
Quantum Logic Array architecture, which was later extended and
improved in~\cite{thaker2006qmh}.  Their work concentrated on
architectural research and did not delve into details of physical
layout or scheduling.  Finally, Metodi et al~\cite{metodi2006spo}
created a tool to automatically generate a physical operations
schedule given a quantum circuit and a fixed grid-based layout
structure.  We extend and improve upon their work by adding new
scheduling heuristics capable of running on grid-based and
non-grid-based layouts.

Maslov et al~\cite{maslov07} have recently proposed heuristics for the
mapping of quantum circuits onto molecules used in liquid state NMR
quantum computing technology.  Their algorithm starts with a molecule
to be used for computation, modeled as a weighted graph with edges
representing atomic couplings within the molecule.  The dataflow graph
of the circuit is mapped onto the molecule graph with an effort to
minimize overall circuit runtime.  Our techniques focus on circuit
placement and routing in an ion trap technology and do not use a
predefined physical substrate topology as in the NMR case.  A new
ion trap geometry is instead generated by our toolset for each
circuit.

%% file: proposedcad.tex
\section{Quantum CAD Flow}\label{sec:proposedcad}

The ultimate goal of a quantum CAD flow is identical to that of a
standard classical CAD flow: to automate the synthesis and laying out
of a circuit.  For a quantum CAD flow, the output circuit consists of
both the quantum portion and the associated classical control
logic.

\begin{figure}
\begin{center}
\epsfig{file=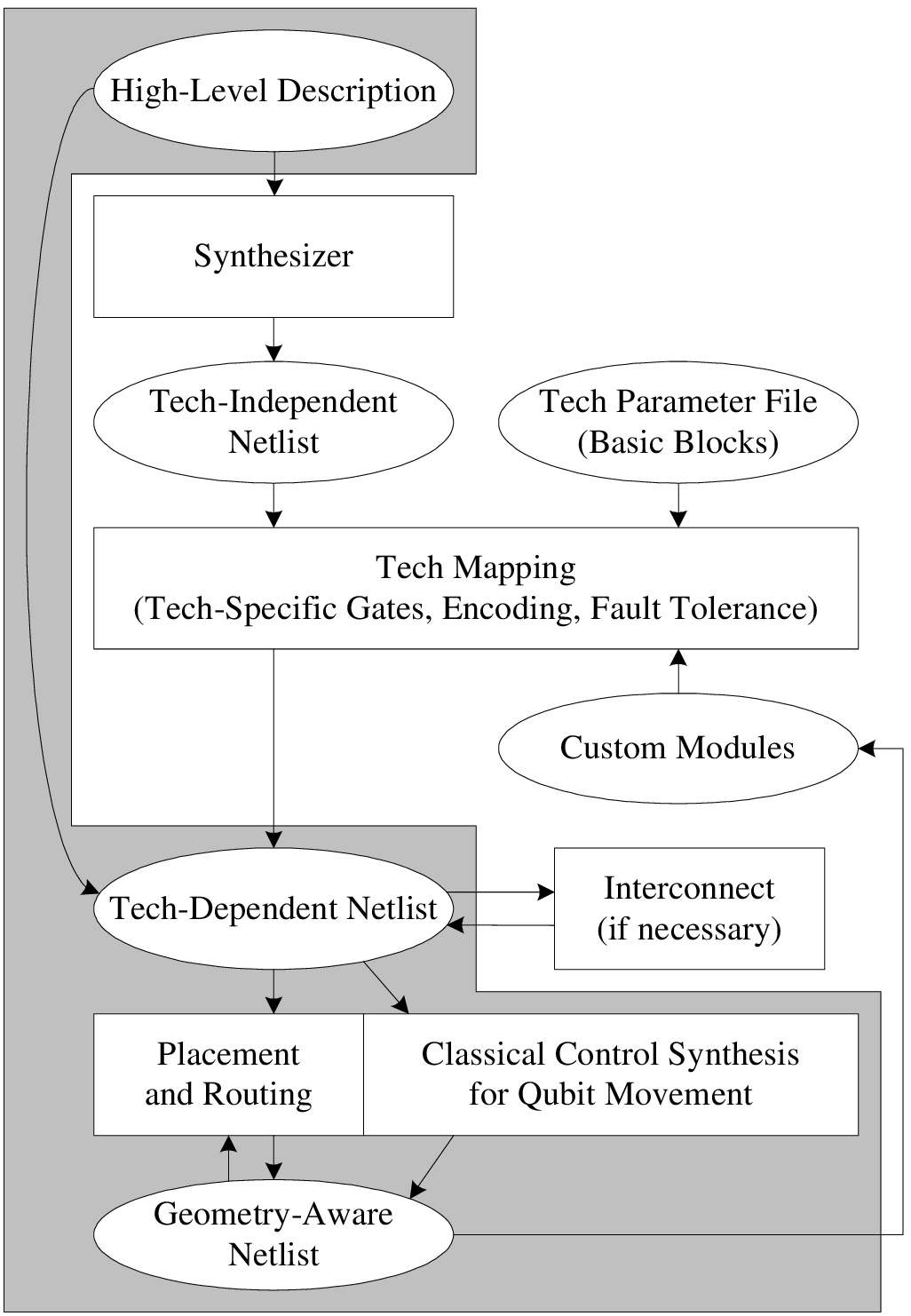,width=0.9\hsize,height=3.5in}
\end{center}
\vspace{-0.2in}
\caption{\label{fig:cadflow}An overview of our CAD flow for quantum circuits.
Ovals represent files; rectangles represent tools.  The gray area highlights
the portions on which we focus in this paper.}
\end{figure}

The quantum CAD flow we present elaborates on the design flows
described in prior works~\cite{balensiefer2005efa, svore2006lsa,
svore04tsa}.  Unlike prior work, our CAD flow addresses the need to
integrate automatic generation of classical control into the flow.
Figure~\ref{fig:cadflow} shows an overview of our CAD toolset.
Rectangles are tools, while ovals represent intermediate file formats.
Our toolset is built to be as similar to classical CAD flows as
possible, while still accounting
for the differences between classical and quantum computing
described in Section~\ref{sec:quantvclass}.

At the top, we begin with a high-level description of the desired quantum circuit.
At present this specification consists of a sequence of quantum assembly
language (QASM~\cite{balensiefer2005efa}) instructions implementing the desired circuit,
since this is a convenient format already being used by various third-party
tools.  We are currently investigating extension of this high-level
description to other formats, such as schematic entry, mathematical
formulae or a more general high-level language.

The synthesizer parses the QASM file and generates a technology-independent
netlist stored in XML format.  From this point onward (downward in the
figure), all file formats are XML.  Additionally, information may be modified
or added but generally not removed.  As we move down the flow, we add more
and more low level details, but we also keep high-level information such as
encoded qubit groupings, nested layout modules, distinction between ancillae
and data, etc.  This allows low-level tools to make more intelligent decisions
concerning qubit placement and channel needs based on high-level circuit
structure.  It likewise allows logical level modification at the lowest
levels without having to attempt to deduce qubit groupings.

A technology parameter file specifies the complete set of basic blocks
available for the layout (see examples in Figure~\ref{fig:macroblocks}),
as well as design rules for connecting them.  A basic block specification
contains the following:
\begin{itemize}
\item the geometry of the block in enough detail to allow fabrication
\item control logic for each operation possible within the block (including
both movement and gates)
\item control logic for handling each operation possible at each interface
\end{itemize}

The most basic function of the technology mapping tool is to take a
technology-independent netlist and map it onto allowed basic blocks to
create the technology-dependent netlist.  This may be more or less
complicated depending upon the complexity of the basic blocks.  In
addition, it may need to translate to technology-specific gates (in case
the QASM file uses gates not available in this technology), encode the
qubits used in the circuit (perhaps also automatically adding
the ancilla and operation sequences necessary for error correction) and
add fault tolerance to the final physical circuit.

In the initial technology-dependent netlist, all qubits are physical
qubits, meaning that encoding levels have been set (though they may still
be modified later).  At this point, any technology-specific
optimizations may optionally be applied to the physical circuit
encapsulated in this netlist.  Additionally, if the circuit is complex
enough to warrant the inclusion of a teleportation-based interconnection
network~\cite{isailovic2006ins}, it is added to the netlist here using the higher
level qubit grouping information in the netlist.

Once the designer is happy with the netlist, a placement and routing
tool lays out the netlist and adds any further channels needed for
communication.  This geometry-aware netlist may be iterated upon as
necessary to refine the layout.  Once the layout is finalized, the classical control
synthesis tool combines the control logic of the various components of
the design, integrates interface control mechanisms to function
properly and generates the unified control structure for the entire
layout.  Our control synthesis tool generates a Verilog
file, which may then be run through a classical CAD flow for
implementation.

The layout specification along with the control logic file together
comprise the geometry-aware netlist, which is the end result for the
quantum circuit initially specified in the high-level description.  In
order to allow hierarchical design of larger quantum circuits, we may
now add this geometry-aware netlist to our set of custom modules.
Future technology mappings may use both the basic blocks specified in
the technology parameter file and any custom modules we create (or
acquire).

The gray area in Figure~\ref{fig:cadflow} identifies the portions we
shall be focusing on for the rest of this paper.  We currently process
the high-level description (a QASM file) directly into a technology-dependent
netlist for ion traps using the macroblocks shown in Figure~\ref{fig:macroblocks}.
Thus we perform a tech mapping, but no automatic encoding, interconnect or
addition of gates for fault tolerance.  In this paper, we focus on laying out
low-level circuits, such as those for encoded ancilla generation and error
correction.  The classical control synthesis box of the CAD flow is discussed
in Section~\ref{sec:control}, while placement and routing are analyzed and
compared in Sections~\ref{sec:grid_layouts},~\ref{sec:greedy_layout},
\ref{sec:dataflow_layout} and~\ref{sec:evaluation}.

We use two main metrics to evaluate the performance of our CAD flow: area
and latency.
For area, we consider the bounding box around the layout, so
irregularly-shaped layouts are penalized (since they have wasted
space).  To determine latency of circuit execution, we use the
scheduling heuristic described in Section~\ref{sec:scheduling} and
extended in Section~\ref{sec:annotated_sched}.  A third metric of
interest is fault-tolerance.  For small layouts and circuits, we can
use third-party tools to determine whether a given layout and schedule
is fault-tolerant~\cite{cross2006qasm}, but we do not currently use
the fault-tolerance metric in our iterative design flow.  We use area
and latency because, to a first approximation, lower area and lower
latency are likely to decrease decoherence.
Previous algorithms to accurately determine the error tolerance of a
quantum circuit have involved very computationally-intensive analyses
that would be inappropriate for circuits with more
than a few dozen gates~\cite{aliferis2005qat}.  However, we are looking
into ways to incorporate fault tolerance as a metric.

%% file: control.tex
\section{Control}\label{sec:control}

The classical control system is responsible for executing the quantum
circuit, including deciding where and when gate operations occur and
tracking and managing every qubit in the system.  It is composed of
the following major components: instruction issue logic, gate control
logic and macroblock control logic.  Instruction issue
logic handles all instruction scheduling and determines qubit movement
paths.  Gate control logic oversees laser resource arbitration,
deciding which requested gate operations may occur at any given time. The
macroblock control logic, which consists of an individual logic block
for each macroblock in the system, handles all the internals of the
macroblock, including details of gate operation for each
gate possible within the macroblock, qubit movement within the
macroblock and qubit movement into and out of the ports.

\subsection{Control Interfaces}\label{sec:control_interface}

\begin{figure}
\begin{center}
\epsfig{file=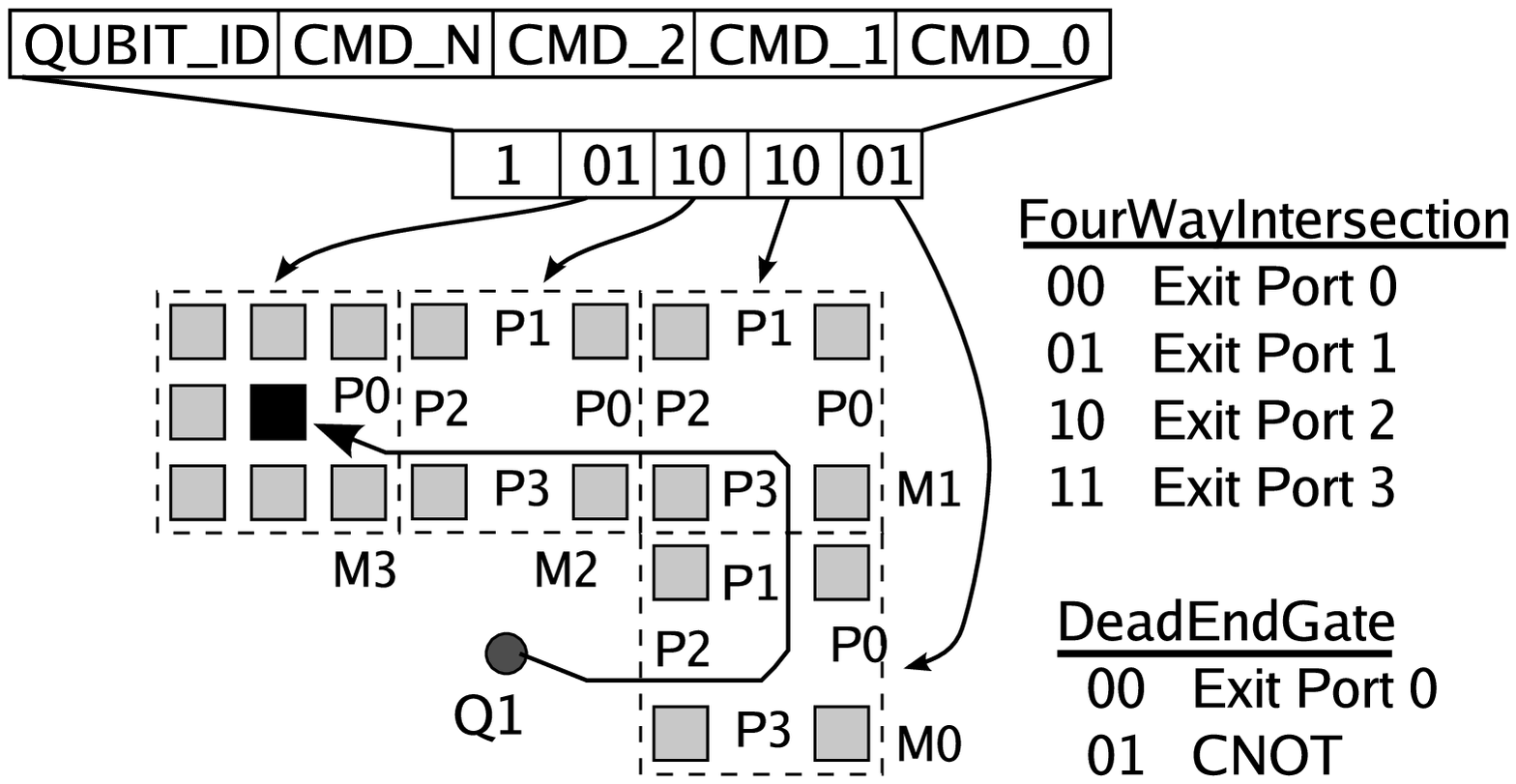,width=\hsize}
\end{center}
\vspace{-0.2in}
\caption{\label{fig:qubit_command_msg}Example of how a qubit control message
is constructed to move a qubit through a series of macroblocks.  The qubit
enters M0 and travels through M1 and M2, arriving at M3 where it is instructed
to perform a CNOT.}
\end{figure}

The first step in the control flow involves processing the quantum
circuit's high-level description (the QASM file).
The instruction issue logic accepts this stream of instructions as input
and creates a series of qubit control messages.  Using these qubit control
messages, macroblock control logic blocks can determine where to move
qubits and when to execute a gate operation.  Qubit control messages
are simple bit streams composed of a qubit ID, along with a sequence
of commands, as shown in Figure~\ref{fig:qubit_command_msg}.  When a
qubit needs to perform an action, the instruction issue logic sends to it
an appropriate control message which travels with the qubit as it traverses
the layout.  Once a macroblock receives a qubit and its
corresponding control message, it uses the first command in the
sequence to determine the operation it must perform.  The macroblock
then removes the command bits used and passes on the remaining
control message to the next macroblock into which the qubit travels.
In this manner, the instruction issue logic can create a multi-command
qubit control message that specifies the path a qubit will traverse
through consecutive macroblocks, along with where gate operations take
place.  The instruction issue logic only has to transmit this control
message to the source macroblock, relying on the inter-macroblock
communication interface to handle the rest.

Communication between the instruction issue logic and the macroblocks
takes place using a shared control message bus in order to minimize
the number of wire connections required by the instruction issue
logic.  Each macroblock listens to the control message bus for
messages addressed to it and only processes messages with a
destination ID that match the macroblock's ID.  A macroblock is
only responsible for monitoring the control message bus if it
contains a qubit that has no remaining command bits.  This condition
generally occurs after a gate operation, when the
instruction issue logic is deciding what action the qubit should take next.
Once the instruction issue logic sends a new control message for the
qubit, the macroblock resumes operation.

Macroblocks communicate with each other via control signals associated
with each quantum port in the macroblock.
Each port has signals to control
qubit movement into the macroblock and signals to control movement
out of the macroblock via that port.  These signals are connected to
the corresponding signals of the neighboring macroblocks.  The
macroblocks assert a \texttt{request} signal to a destination macroblock when a
qubit command indicates the qubit should cross into the next
macroblock.  If an \texttt{available} signal response is received, the qubit,
along with its control message, can move across into the neighboring
macroblock; if not, the qubit must wait until the available signal is
present.

The macroblock interface enables the instruction issue logic to
schedule qubit movement as a path through a sequence of macroblocks,
without concerning itself with the low level details of qubit
movement.  This modular system allows macroblocks to be replaced with
any other macroblock that implements the defined interface, without
modifying the instruction issue logic.

Additionally, macroblocks have an interface to the laser control
logic.  Whenever a macroblock is instructed to perform a gate operation,
it must request a laser resource through the laser control logic.
The laser controller is responsible for aggregating requests from
all the macroblocks in the system, and deciding when and where to
send laser pulses.  The laser controller also attempts to parallelize
as many operations as possible.  Once the laser pulses have completed,
the laser controller notifies the macroblocks, indicating that the gate
operation is complete.

\subsection{Instruction Scheduling}\label{sec:scheduling}

The instruction issue logic is responsible for determining the runtime
execution order of the instructions in the quantum circuit, which involves
both preprocessing and online scheduling.  The instruction
sequence is first preprocessed to assign priorities that will help during
scheduling.  The sequence is traversed from end to beginning, scheduling
instructions as late as dependencies allow, using realistic gate latencies
but ignoring movement.  Essentially, each instruction is labeled with the
length of its critical path to the end of the program.  This is similar to
the method used in~\cite{metodi2006spo}, but we use critical path with gate
times rather than the size of the dependent subtree.

The instruction preprocessing generates an optimal schedule assuming
infinite gates and zero movement cost.  However, we wish to evaluate a
layout with more realistic characteristics.  Our scheduler is designed to schedule on an arbitrary
graph, but the layouts provided to it by the place and route tool
are in fact planar layouts using only right angles.  In addition, the scheduler
requires that the qubit initial positions be provided as well.

Our scheduler implements a greedy scheduling technique.  It keeps the set
of instructions which have had all their dependencies fulfilled (and thus
are ready to be executed).  It attempts to schedule them in priority order.
So the highest priority ready instruction (according to critical path) is
attempted first and is thus more likely to get access to the resources it needs.
These contested resources include both gates and channels/intersections.
Once all possible instructions have been scheduled, time advances until
one or more resources is freed and more instructions may be scheduled.
This scheduling and stalling cycle continues until the full sequence has
been executed or until deadlock occurs, in which case it is detected and
the highest priority unscheduled instruction at the time of deadlock is
reported.

Since we are interested in evaluating layouts rather than in designing an
efficient online scheduler, we use very thorough searches over the graph in both gate
assignment and pathfinding.  This causes the scheduler to take longer but
takes much of the uncertainty concerning schedule quality out of our tests.
In addition, the scheduler reports stalling information which may be used
for iterating upon the layout.

\subsection{Control Extraction}

Armed with well defined component interfaces and a method to execute
the quantum instructions, all that remains to create the control
system for a given quantum circuit is putting the pieces together.
The quantum datapath is composed of an arbitrary number of macroblocks
pulled from the component library.  Each macroblock in our component
library has associated with it classical control logic.  The control
logic handles all the internals of the macroblock including details of
ion movement, ion trapping and gate operation.  In our library,
the macroblock control logic is specified using behavioral Verilog
modules.

When the layout stage of the CAD flow creates a physical layout of
macroblocks, we extract the corresponding control logic blocks and
assemble them together in a top-level Verilog module for the full
control system, stitching together all necessary macroblock interfaces.
This module instantiates all the appropriate
macroblock control modules, along with the instruction issue logic and
laser controller unit.  Combined, these modules are assembled into a
single Verilog module which implements the full classical control system
for the quantum circuit and which may be input to a classical CAD flow
for synthesis.

%% file: grid_layout.tex

\begin{figure}
\begin{center}
\epsfig{file=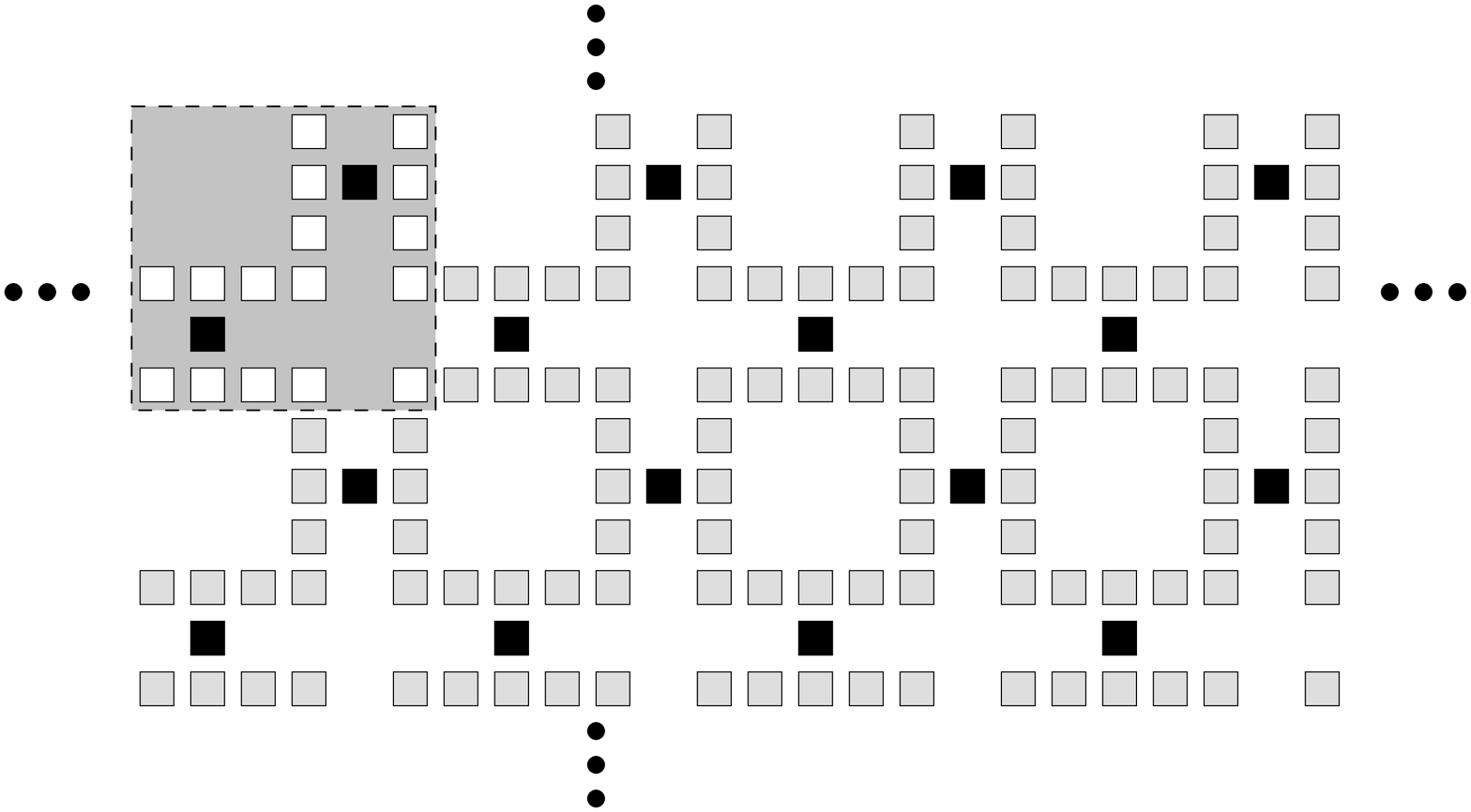, width=.9\hsize}
\end{center}
\vspace{-0.2in}
\caption{\label{fig:setso_grid}QPOS grid structure constructed
by tiling the highlighted $2\times 2$ macroblock cell.}
\end{figure}

\begin{figure*}
\begin{minipage}[b]{\columnwidth}
\begin{center}
\epsfig{file=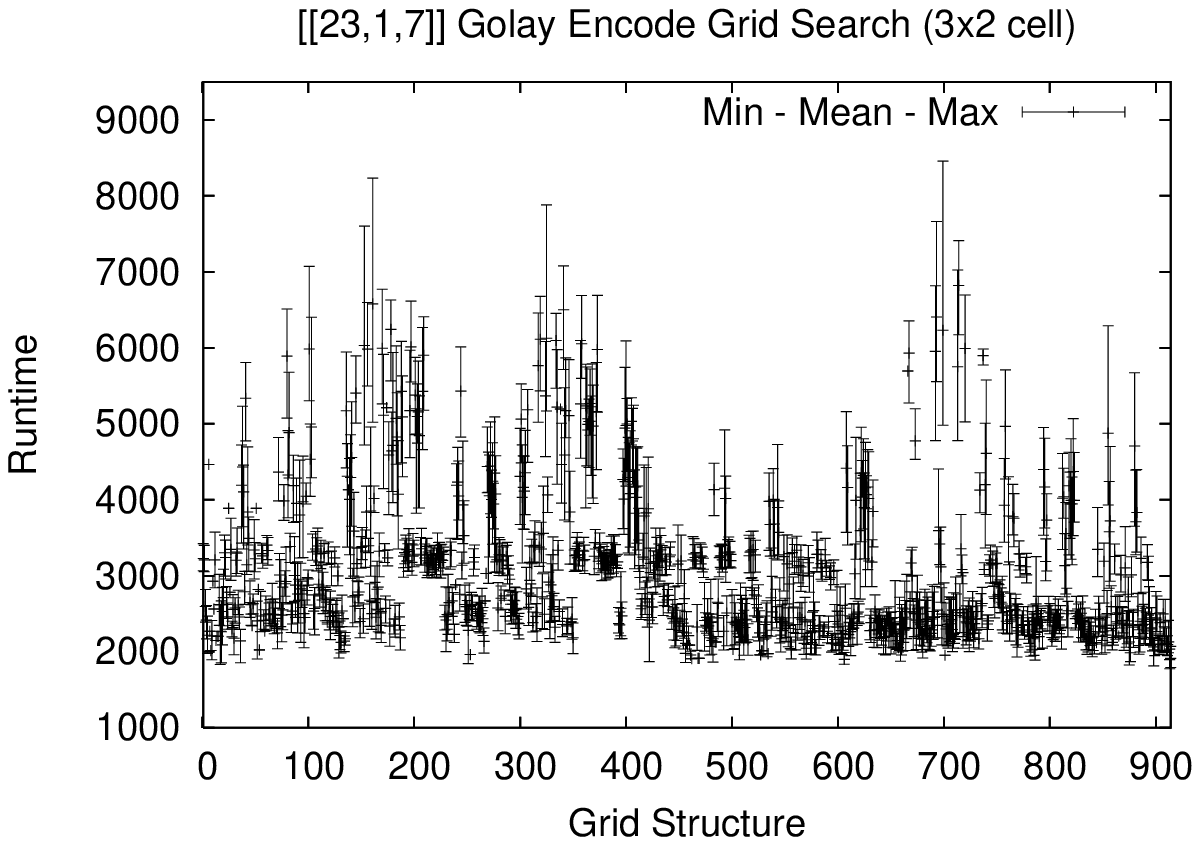, width=\hsize}
\end{center}
\end{minipage}\hfill
\begin{minipage}[b]{\columnwidth}
\begin{center}
\epsfig{file=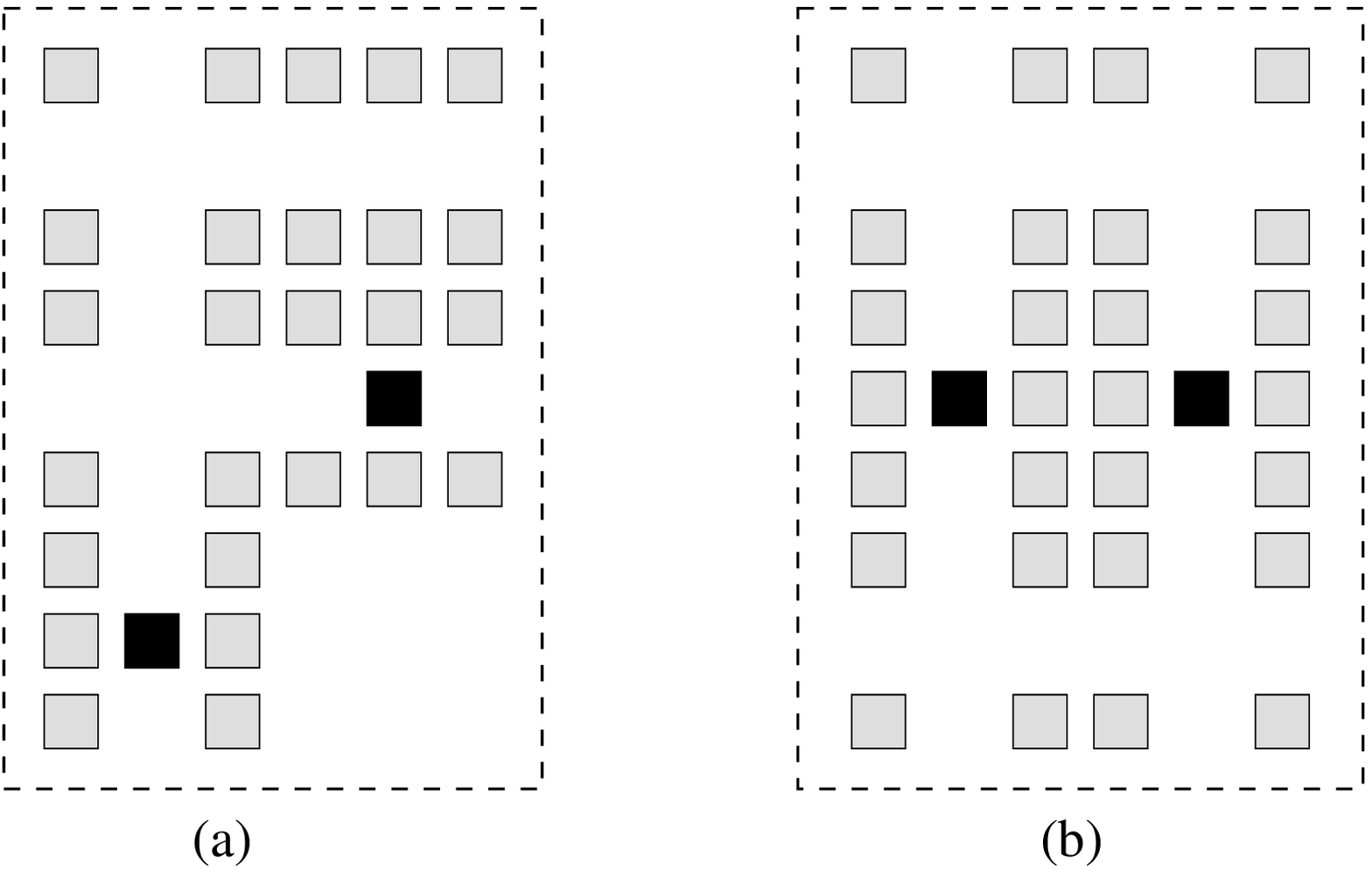, width=\hsize}
\end{center}
\end{minipage}
\hfill
\begin{minipage}[t]{\columnwidth}
\caption{\label{fig:23_1_7_golay_encode_level1_search_143_3by2}Variations in runtime
of various grid-based physical layouts for  $[[23,1,7]]$ Golay encode circuit. For each
grid structure the minimum, mean, and maximum time are plotted. }
\end{minipage}\hfill
\begin{minipage}[t]{\columnwidth}
\caption{\label{fig:grid_structures}Comparison of the best $3\times 2$ cell
for two different circuits.  (a) The best cell for the $[[23,1,7]]$ Golay encode
circuit.  (b) The best cell for the $[[7,1,3]]$ L1 correct circuit.}
\end{minipage}
\end{figure*}

\section{Grid-based Layouts}\label{sec:grid_layouts}

We begin our exploration of placement and routing heuristics by
considering grid-based layouts.  A majority of the work done in the
field has concentrated on these types of
layouts.
In all of these works, a layout is constructed by first designing a
primitive cell and then tiling this cell into a larger physical layout.
For example, the authors of \cite{metodi2005qla, metodi2006spo}
manually design a single cell, and for any given quantum circuit, they
use that cell to construct an appropriately sized layout.  In
\cite{svore04tsa}, the authors automate the generation of an H-Tree
based layout constructed from a single cell pattern.  Similarly,
\cite{balensiefer2005efa} uses a cell such as in \cite{svore04tsa} but
also provides some tools to evaluate the performance of a circuit when
the number of communication channels and gate locations within the
cell is varied.  We use a combination of these methods to implement a
tool that automatically creates a grid-based physical layout for a
given quantum circuit.

The grid-based physical layouts generated by our tools are constructed
by first creating a primitive cell out of the macroblocks mentioned in
Section~\ref{sec:iontraps} and then tiling the cell to fill up the desired
area.  For example, Figure~\ref{fig:setso_grid} shows how a $2\times
2$ sized cell can be tiled to create the layout used in
\cite{metodi2006spo} (referred to henceforth as the QPOS grid).  These types of
simple structures are easy to automatically generate given only the
number of qubits and gate operations in the quantum circuit.  Furthermore,
grid-based structures are very appealing to consider because, apart from
selecting the number of cells in the layout and the initial qubit
placement, no other customization is required in order to map a
quantum circuit onto the layout.  The regular pattern also
makes it easy to determine how qubits move through the system, as
simple schemes such as dimension-ordered routing can be
used.

The approach we use to generate the grid-based layout for a given
quantum circuit is as follows:
\begin{enumerate}
\item Given the cell size, create a valid cell structure out of
  macroblocks.
\vspace{-4pt}
\item Create a layout by tiling the cell to fill up the desired area.
\vspace{-4pt}
\item Assign initial qubit locations.
\vspace{-4pt}
\item Simulate the quantum circuit on the layout to determine the
  execution time.
\end{enumerate}

The first step finds a valid cell structure.  A cell is valid
if all the macroblocks that open to the perimeter of the cell have
an open macroblock to connect to when the cell is tiled.  Also, a
cell cannot have an isolated macroblock within it that is 
unreachable.  Once we tile this valid cell to create a larger
layout, we must decide on how to assign initial qubit locations.
The two methods we utilize are: a systematic
left to right, one qubit per cell approach, and a randomized
placement.  The systematic placement allows us to fairly compare
different layouts.  However, since the initial placement of the qubits
can affect the performance of the circuit, the tool also tries a number
of random placements in an effort to determine if the systematic
placement unfairly handicapped the circuit.

This layout generation and evaluation procedure is iterated upon until
all valid cell configurations of the given size are searched.  We then
repeat this process for different cell sizes.  The cell structure that
results in the minimum simulated time for the circuit is used to
create the final layout.

As an example,
Figure~\ref{fig:23_1_7_golay_encode_level1_search_143_3by2} shows the
results of searching for the best layout composed of $3\times 2$ sized
cells targeting the $[[23,1,7]]$ Golay encode circuit~\cite{steane2003oan},
one of our benchmarks shown in Table~\ref{table:circuits}.
More than 900 valid cell
configurations were tested.  For each cell configuration, we try
multiple initial qubit placements (as mentioned earlier) resulting in
a range of runtimes for each cell configuration. Differences in the
runtime of the circuit are not limited to just variations on the cell
configuration but are in fact also highly dependent on the initial
qubit placement.

Figure~\ref{fig:grid_structures} shows the best cell structure found
by conducting a search of all $2\times 2$, $2\times 3$, and $3\times
2$ sized cells for two different circuits. The main result of this
search is that the best cell structure used to create the grid-based
layout is dependent on  what circuit will be run upon it.  By varying
the location of gates and communication channels, we tailor
the structure of the layout to match the circuit requirements.

While this type of exhaustive search of physical layouts is capable of
finding an optimal layout for a quantum circuit, it suffers from a
number of drawbacks.  Namely, as the size of the cell increases, the
number of possible cell configurations grows exponentially.  Searching
for a good layout for anything but the smallest cell sizes is not a
realistic option.  Furthermore, while small circuits may be able to
take advantage of primitive cell based grids, larger circuits will require
a less homogeneous layout.  One approach to doing this is to construct
a large layout out of smaller grid-based pieces, all with different
cell configurations.  
While this approach is interesting, we feel a
more promising approach is one that resembles a classical CAD
flow, where information extracted from the circuit is used to construct
the layout.

%% file: greedy_layout.tex
\section{Greedy Place and Route}\label{sec:greedy_layout}

One problem we observed in the regular grid layout design was that the
high amount of channel congestion due to limited bandwidth
causes densely-packed (occupied) gates.  Additionally, we found that a
number of gate locations and channels in many of the grids were not
even used by the scheduler to perform the circuit.  

We present a new heuristic that attempts to solve some of
these problems.  The heuristic is a simple greedy algorithm that
starts with only as many gate locations as qubits (because we assume
that qubits only rest in storage/gate locations) and no channels
connecting the gates.  It iterates with the circuit scheduler, moving
and connecting gate locations until the qubits can communicate
sufficiently to perform the specified circuit.  The current layout is
fed into the circuit scheduler which tries to schedule until it finds
qubits in gate locations that cannot communicate to
perform a gate.  The place and router then connects the problematic
gate locations and tries scheduling on the layout again.  The iteration
finishes once the circuit can be successfully completed.
Our algorithm bears some similarity
to the iterative procedure in adaptive cluster growth
placement~\cite{kyung1990acg} in classical CAD.
Gate locations are placed from the center outward as the
circuit grows to fit a rectilinear boundary.

The placer can move gate locations that have to be connected if they
are not already connected to something else.  The router connects gate
locations by making a direct path in the x and y directions between
them and placing a new channel, shifting existing channels out of
the way. Since channels are allowed to overlap, intersections are inserted
where the new channels cut across existing ones.

This technique has the advantage that, since the circuit scheduler
prioritizes gates based on gate delay critical path, potentially
critical gates are mapped to gate locations and connected early in the
process.  Thus critical gates tend to be initially placed close
together to shorten the circuit critical path.  Additionally, gate
locations that need to communicate can be connected directly instead
of using a general shared grid channel network, where congestion can
occur and cause qubits to be routed along unnecessarily long paths.

A disadvantage of this heuristic is that gate placement is done to
optimize critical path, not to minimize channel intersections.  This
means that the layout could end up having many 4-way channel intersections
and turns, both of which have more delay than
2-way straight channels.  Additionally, even though critical gates are
mapped and placed near each other, the channel routing algorithm tends
to spread these gate locations apart as more channels cut through the
center of the circuit.  We discuss our experimental evaluation of this
heuristic in Section~\ref{sec:evaluation}.

%% file: dataflow_layout.tex
\section{Dataflow-Based Layouts}\label{sec:dataflow_layout}

As described in Section~\ref{sec:grid_layouts}, a systematic row by row initial
placement for qubits allows us to make somewhat accurate comparisons between
different grid-based layouts, while a random initial qubit placement allows us
to test a single grid's dependence on qubit starting positions.  However, in
laying out a quantum circuit, we would like to have a more intelligent and
natural means of determining initial qubit placement.  For this, we turn to
the dataflow graph representation of the circuit.

\subsection{Dataflow Graph Analysis}\label{sec:graph_analysis}

Figure~\ref{fig:dataflow_graph}a shows a QASM instruction sequence consisting
of Hadamard gates (H) and controlled bit-flips (CX) operating on qubits Q0, Q1,
Q2 and Q3, with each instruction labeled by a letter.  Figure~\ref{fig:dataflow_graph}b
shows the equivalent sequence of operations in standard quantum circuit
format.  Either of these may be translated into the dataflow graph
shown in Figure~\ref{fig:dataflow_graph}c, where each node represents a QASM
instruction (as labeled in Figure~\ref{fig:dataflow_graph}a) and each arc
represents a qubit dependency.
With this dataflow graph, we may perform
some analyses to help us place and route a layout for our quantum circuit.

The general idea is that we shall create node groups in the dataflow graph
which correspond to distinct gate locations that may then be placed and routed
on a layout.  All instructions within a single node group are guaranteed to be
executed at a single gate location, as elaborated upon in Section~\ref{sec:annotated_sched}.
To begin with, we create a node group for each instruction, giving us a dataflow
group graph, as shown in Figure~\ref{fig:dataflow_alg}a.  If we lay out
this group graph with a distinct designated gate for each instruction (using
heuristics discussed in Section~\ref{sec:pnr}), we get a layout in which
the starting location of each qubit is specified implicitly by its first gate
location, so no additional initial placement heuristic is needed.

From this layout we can extract movement latency between nodes and label the edges
with weights (as in Figure~\ref{fig:dataflow_alg}a).  We now find the longest
critical path by qubit.  The critical path A-E-I of qubit Q0 has length 14 (the
dashed bold arrows), while the critical path C-F-G-H-I of qubit Q2 has length 15
(the solid bold arrows).  We select the longest edge on the longest critical path,
which is the edge G-H with weight 5.  We merge these two node groups to eliminate
this latency, in effect specifying that these two instructions should occur
at the same gate location (Figure~\ref{fig:dataflow_alg}b).  We then update the layout and recompute distances.
Assuming we merged these two node groups to the location of H (NG8), then the
weight of edge F-G changes to 1 (to match the weight of edge F-H) and the weight
of edge E-G probably changes to 6 (former E-G plus former G-H), but the exact change really
depends on layout decisions.  The new critical path is now A-E-I, so if we
do this again, we merge node groups NG5 and NG9 to eliminate the edge of weight 8,
and we get the group graph in Figure~\ref{fig:dataflow_alg}c.

In merging nodes, there is the possibility that two qubit starting locations get
merged, complicating the assignment of initial placement.  For this reason, we
add a dummy {\it input} node for each qubit before its first instruction.  The
merging heuristic doesn't allow more than one input node in any single node group,
so we maintain the benefit of having an intelligent initial qubit placement without
extra work.

There is an important trade-off to consider when taking this merging approach.
A tiled grid layout provides plenty of gate location reuse but is unlikely to
provide any pipelinability without great effort.  A layout of the group graph in
Figure~\ref{fig:dataflow_alg}a (with each instruction assigned to a distinct gate
location) provides no gate location reuse at all but high potential pipelinability.  This
raises the question of whether we wish to minimize area and time (for critical
data qubits), maximize throughput of a pipeline (for ancilla generation), or
compromise at some middle ground where small sets of
nearby nodes are merged in order to exploit locality while still retaining
some pipelinability.  We intend to further explore this topic in the future.

\subsection{Placement and Routing}\label{sec:pnr}

Taking the group graph from the dataflow analysis heuristic, the
placement algorithm takes advantage of the fanout-limited gate output
imposed by the No-Cloning Theorem~\cite{nocloning1982wootters} to lay out the dataflow-ordered gate
locations in a roughly rectangular block.  We adopt a gate array-style
design, where gate locations are laid out in columns according to the
graph, with space left between each pair of columns for necessary channels.  This
can lead to wasted space due to a linear layout of uneven column sizes,
so we may also perform a folding operation, wherein a short column may
be folded in (joined) with the previous column, thus filling out the
rectangular bounding box of the layout as much as possible and decreasing area.
The columns are then sorted to position gate locations that
need to be connected roughly horizontal to one another.  This further
minimizes channel distance between connected gate locations and
reduces the number of high-latency turns.

Once gate locations are placed, we use a grid-based model in which we
first route local wire channels between gate locations that are in
adjacent or the same columns.  These channels tend to be only a few
macroblocks long each.  A separate global channel is then inserted
between each pair of rows and between each pair of columns of gate
locations.  These global channels stretch the full length of the layout.
There are no real routing constraints in our simple model
since channels are allowed to overlap and turn into 3- or 4-way
intersections.  We depend on the dataflow column sorting in the
placement phase to reduce the number of intersections and shared
local channels.  While local channels could technically be used for
global routing and vice versa, we've found that this division in
routing tends to divide the traffic and separate local from
long-distance congestion.

\begin{figure}
\begin{center}
\epsfig{file=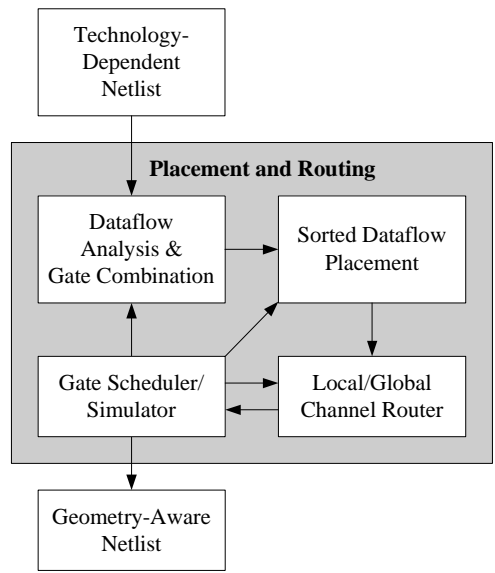, width=0.8\hsize}
\end{center}
\vspace{-0.1in}
\caption{\label{fig:placeroute_flow}The placement and routing portion
  of our CAD flow (shown in Figure~\ref{fig:cadflow}) takes a
  technology-dependent netlist and translates it into a geometry-aware
  netlist through an iterative process involving dataflow analysis and
  placement and routing techniques.}
\end{figure}

With these basic placement and routing schemes, we may now iterate
upon the layout, as shown in Figure~\ref{fig:placeroute_flow}.  The
technology-dependent netlist is translated into a dataflow group
graph with a separate gate location for each instruction
(Figure~\ref{fig:dataflow_alg}a).  This group graph is then placed,
routed and scheduled to get latency and identify the runtime critical
path (as opposed to the critical path in the group graph, which
fails to take congestion into account).  The longest latency move
on the runtime critical path (between two node groups) is merged
into one node group, thus eliminating the move since a node group
represents a single gate location.  This new group graph is then
placed, routed and scheduled again to find the next pair of node
groups to merge.

Once this process has iterated enough times, we reach a point where
congestion at some heavily merged node group is actually hurting the
latency with each further merge.  We alleviate this congestion
by adding storage nodes (essentially gate locations that don't
actually perform gates) near the congested node group.  This increases
the area slightly but maintains the locality exploited by the
merging heuristic.  If congestion persists, we halt the algorithm,
back up a few merging steps and output the geometry-aware netlist.

\subsection{Annotated Scheduling}\label{sec:annotated_sched}

The scheduling heuristic described in Section~\ref{sec:scheduling} schedules an arbitrary
QASM instruction sequence on an arbitrary layout.  However, once we have assigned
instructions in a dataflow graph to node groups (as described in
Section~\ref{sec:graph_analysis}), we wish those instructions to be executed at
their proper location on any layout placed and routed from the group graph.
To this end, we annotate each instruction in the instruction sequence with the
name of the gate location where it must be executed.  Additionally, since we
have the gate locations in advance, we can incorporate movement in the
back-prioritization of the instruction sequence.  Thus, the priority assigned to
each qubit now incorporates both gate latencies and movement through an uncongested
layout, which gives us a better approximation of each qubit's critical
path.  We use this extended
scheduler in our dataflow-based experiments presented in Section~\ref{sec:evaluation}.

%% file: evaluation.tex
\section{Results}\label{sec:evaluation}

We now present our simulation results for the heuristics described
in earlier sections.

\subsection{Benchmarks}

\begin{table}
\begin{center}
  \begin{tabular}{|l|r|r|}
    \hline
     & Qubit & Gate \\
    Circuit name & count & count \\
    \hline \hline
        $[[7,1,3]]$ L1 encode~\cite{Ste96c} & 7 & 21 \\
        $[[23,1,7]]$ L1 encode~\cite{steane2003oan} & 23 & 116 \\
        $[[7,1,3]]$ L1 correction~\cite{aliferis2005qat} & 21 & 136 \\
        $[[7,1,3]]$ L2 encode~\cite{Ste96c} & 49 & 245 \\
        \hline
\end{tabular}
\end{center}
\vspace{-0.2in}
\caption{List of our QECC benchmarks, with quantum gate count and
  number of qubits processed in the circuit.}
\label{table:circuits}
\end{table}

\begin{table*}
\begin{center}
  \begin{tabular}{|l|l|r|r|}
    \hline
    Circuit & Heuristic & Latency ($\mu s$) & Area \\
    \hline
    \hline
    $[[7,1,3]]$ L1 encode 
    & QPOS Grid & 548.0 & 49 \\
    & Optimal Grid & 509.0 & 49 \\
    & Greedy channel and gate location placement
    & 648.0 & 36 \\
     & Non-folded DF, 2 global channels, critical combining
    & 768.2 & 231 \\
     & Folded DF, 1 global channels, critical combining
    & 795.4 & 126 \\
     & Folded DF, 2 global channels, critical combining
    & 712.4 & 182 \\
    \hline
    $[[23,1,7]]$ Golay encode 
    & QPOS Grid & 2268.0 & 575 \\
    & Optimal Grid & 1801.0 & 575 \\
    & Greedy channel and gate location
    placement & 2457.0 & 168 \\
     & Non-folded DF, 2 global channels, critical combining
    & 2169.2 & 3880 \\
     & Folded DF, 1 global channels, critical combining
    & 2264.0 & 713 \\
     & Folded DF, 2 global channels, critical combining
    & 2248.2 & 1394 \\
    \hline
    $[[7,1,3]]$ L1 correction 
    & QPOS Grid & 1300.0 & 1271 \\
    & Optimal Grid & 771.0 & 1271 \\
    & Greedy channel and gate location
    placement & 1932.0 & 756 \\
     & Non-folded DF, 2 global channels, critical combining
    & 999.8 & 2378 \\
     & Folded DF, 1 global channels, critical combining
    & 1501.2 & 690 \\    
     & Folded DF, 2 global channels, critical combining
    & 1121.2 & 1496 \\
    \hline
    $[[7,1,3]]$ L2 encode 
    & QPOS Grid & 2411.0 & 1365 \\
    & Optimal Grid & 1367.0 & 1365 \\
    & Greedy channel and gate location placement
    & 4791.0 & 936 \\    
     & Non-folded DF, 2 global channels, critical combining
    & 1582.4 & 4087 \\
     & Folded DF, 1 global channels, critical combining
    & 1828.6 & 1617 \\    
     & Folded DF, 2 global channels, critical combining
    & 1944.8 & 3381 \\
    \hline
  \end{tabular}
\end{center}
  \caption{Latency results for a variety of ECC circuits with
    different placement and routing heuristics.}
  \label{table:latencyHeuristic}
\end{table*}

Relatively high error rates of operations in a quantum computer
necessitate heavy encodings of qubits.  As such, we focus on
encoding circuits (useful for both data and ancillae) and error
correction circuits to experiment with circuit layout techniques.
We lay out a number of error
correction and encoding circuits to evaluate the effectiveness of the
heuristics used in our CAD flow in terms of circuit area and latency,
as determined by our scheduler.  Our circuit benchmarks are shown in
Table~\ref{table:circuits}.  We use two level 1 (L1) encoding circuits,
a level 2 (L2) recursive encoding circuit and a fault-tolerant level 1
correction circuit.

The idea of the encoding circuits is that they will provide a constant
stream of encoded ancillae to interact with encoded data qubit blocks.
Thus, for these circuits, throughput is a more important measure than
latency, implying that they would benefit greatly from pipelining.
Nonetheless, a high latency circuit could introduce non-trivial error
due to increased qubit idle time. 
On the other hand, correction circuits are much more latency dependent,
since they are on the critical path for the processing of data qubit
blocks.

\subsection{Evaluation}

We have evaluated a variety of layout design heuristics on the four benchmarks
shown in Table~\ref{table:circuits}.  The results are in Table~\ref{table:latencyHeuristic}.
``QPOS Grid'' refers to the best scheduled layout from the literature~\cite{metodi2006spo}
(see Section~\ref{sec:grid_layouts}).
``Optimal Grid'' refers to the best grid with an area matching the QPOS Grid used that was found by the exhaustive
search described in Section~\ref{sec:grid_layouts}.
``Greedy'' refers to the heuristic described in Section~\ref{sec:greedy_layout}.
``DF'' refers to the dataflow-based approach from Section~\ref{sec:dataflow_layout}.
``Non-folded'' means the dataflow graph is laid out with varying column widths; ``folded'' means the
layout has been made more rectangular by stacking columns.
The number of global channels is between each pair of rows and columns of gate locations.
``Critical combining'' refers to our dataflow group graph merging heuristic.

The exhaustive search over grids yields the best latency for all benchmarks,
which is not surprising.  This kind of search becomes intractable quickly as
circuit size grows, and additionally, it is based on the unproven assumption
that a regular layout pattern is the best approach.  We include this data point
as something to keep in mind as a target latency.

Among the polynomial-time heuristics, 
we first note that no single heuristic is optimal for all four benchmarks
and that, in fact, no single heuristic optimizes both latency and area for any single
circuit.  Dataflow-based place and route techniques in general produce the
lowest latency circuits.  We find that the optimal global channel count per column (1 or 2)
depends on the circuit being laid out.  This is an artifact of the lack of
maturity in our routing methodology.  We intend to explore more adaptive
routing optimization in our ongoing work.

The dataflow approach and the QPOS Grid tend to trade off between latency and area.
However, we expect that the dataflow approach will show greater potential for
pipelining, thus allowing us to target circuits such as an encoded ancilla generation
factory, for which throughput is of greater importance than latency.  We also
observe that non-folded dataflow layouts are likely to have even greater
pipelinability than folded ones, but at the likely cost of greater area.  Although,
we should note that the area estimates for the non-folded DF-based layouts are in fact overestimates due
to our use of a liberal bounding box for these calculations.

We find that the greedy heuristic tends to find the best design area-wise, but the
latency penalty increases with circuit complexity.  This is expected, as greedy is
unable to handle congestion problems, so it works best for small circuits where
congestion is not an issue.  It is for the opposite reason that the DF heuristics
fail on the $[[7,1,3]]$ L1 encode.  They insert too much complexity into an
otherwise simple problem.

%% file: conclusion.tex
\section{Conclusion}\label{sec:conclusion}

We presented a computer-aided design flow for the layout, scheduling
and control of ion trap-based quantum circuits.  We focused on physical
quantum circuits, that is, ones for which all ancillae, encodings and
interconnect are explicitly specified.  We explored several mechanisms
for generating optimal layouts and schedules for our benchmark circuits.

Prior work has tended to assume a specific regular grid structure and to
schedule operations within this structure.  We investigated a variety of
grid structures and showed a performance variance of a factor of four as
we varied grid structure and initial qubit placement.  Since exhaustive
search is clearly impractical for large circuits, we also explored two
polynomial-time heuristics for automated layout design.  Our \emph{greedy
algorithm} produces good results for very simple circuits, but quickly
begins to be suboptimal as circuit size grows.  For larger circuits, we
investigated a \emph{dataflow-based analysis} of the quantum circuit to
assist a place and route mechanism which leverages from classical
algorithms.  We found that our our dataflow approach generally offers
the best latency, often at the cost of area.  However, we expect that a
layout based on the dataflow graph analysis also offers better potential
for pipelining than a grid-based approach, and we intend to investigate
this further in the future.